\documentclass[aps,prl,twocolumn,superscriptaddress]{revtex4}

\usepackage{graphicx}
\usepackage{amsmath}

\begin{document}

\title{First-principles phase-coherent transport in 
metallic nanotubes with realistic contacts}

\author{J.~J.~Palacios}
\affiliation{Departamento de F\'{\i}sica Aplicada, Universidad de
Alicante, San Vicente del Raspeig, Alicante 03690, Spain.}
\author{A. J. P\'erez-Jim\'enez}
\affiliation{Departamento de Qu\'{\i}mica-F\'{\i}sica, Universidad de
Alicante, San Vicente del Raspeig, Alicante 03690, Spain.}
\author{E. Louis}
\affiliation{Departamento de F\'{\i}sica Aplicada, Universidad de
Alicante, San Vicente del Raspeig, Alicante 03690, Spain.}
\author{E. SanFabi\'an}
\affiliation{Departamento de Qu\'{\i}mica-F\'{\i}sica, Universidad de
Alicante, San Vicente del Raspeig, Alicante 03690, Spain.}
\author{J. A. Verg\'es.}
\affiliation{Departamento de Teor\'{\i}a de la Materia Condensada, Instituto
de Ciencias de Materiales de Madrid (CSIC), Cantoblanco, Madrid 28049, Spain.}

\date{\today}

\begin{abstract}
We present first-principles calculations of phase 
coherent electron transport in a carbon
nanotube (CNT) with realistic contacts. 
We focus on the zero-bias response of open metallic CNT's 
considering two archetypal contact geometries (end and side) and 
three commonly used metals as electrodes (Al, Au, 
and Ti). Our {\em ab-initio} electrical transport calculations make,
for the first time, quantitative predictions on the contact transparency and 
the transport properties of finite metallic CNT's.
Al and Au turn out to make poor contacts while Ti is
the best option of the three. Additional information on the CNT band mixing at 
the contacts is also obtained.
\end{abstract}

\pacs{}

\maketitle

Controversy on the observed
electrical transport properties of carbon nanotubes (CNT's) has been
mostly due to our lack of control and understanding 
of their contact to the metallic electrodes. 
It has finally become clear that the contact  
influences critically the overall performance of the CNT and
that it is crucial to lower the inherent 
contact resistance to achieve the definite
understanding of the intrinsic electrical properties of 
CNT's\cite{Frank:science:98,Bachtold:prl:00,Nygard:nature:00}.
In order to determine the relevant factors behind the 
contact resistance so that this can be pushed down
to its alleged quantum limit $R_0=h/2e^2$ per CNT channel a big
experimental effort has been made both in CNT growth and lithographic 
techniques\cite{Soh:apl:99,Zhou:prl:00,Appenzeller:apl:01,Kong:prl:01,Kanda:apl:01,Liang:prl:02,Derycke:apl:02}. 
While considerable progress in this direction has already been achieved,
theoretical progress, on the other hand, lags behind in this important issue.

\begin{figure}
\includegraphics[width=2.0in]{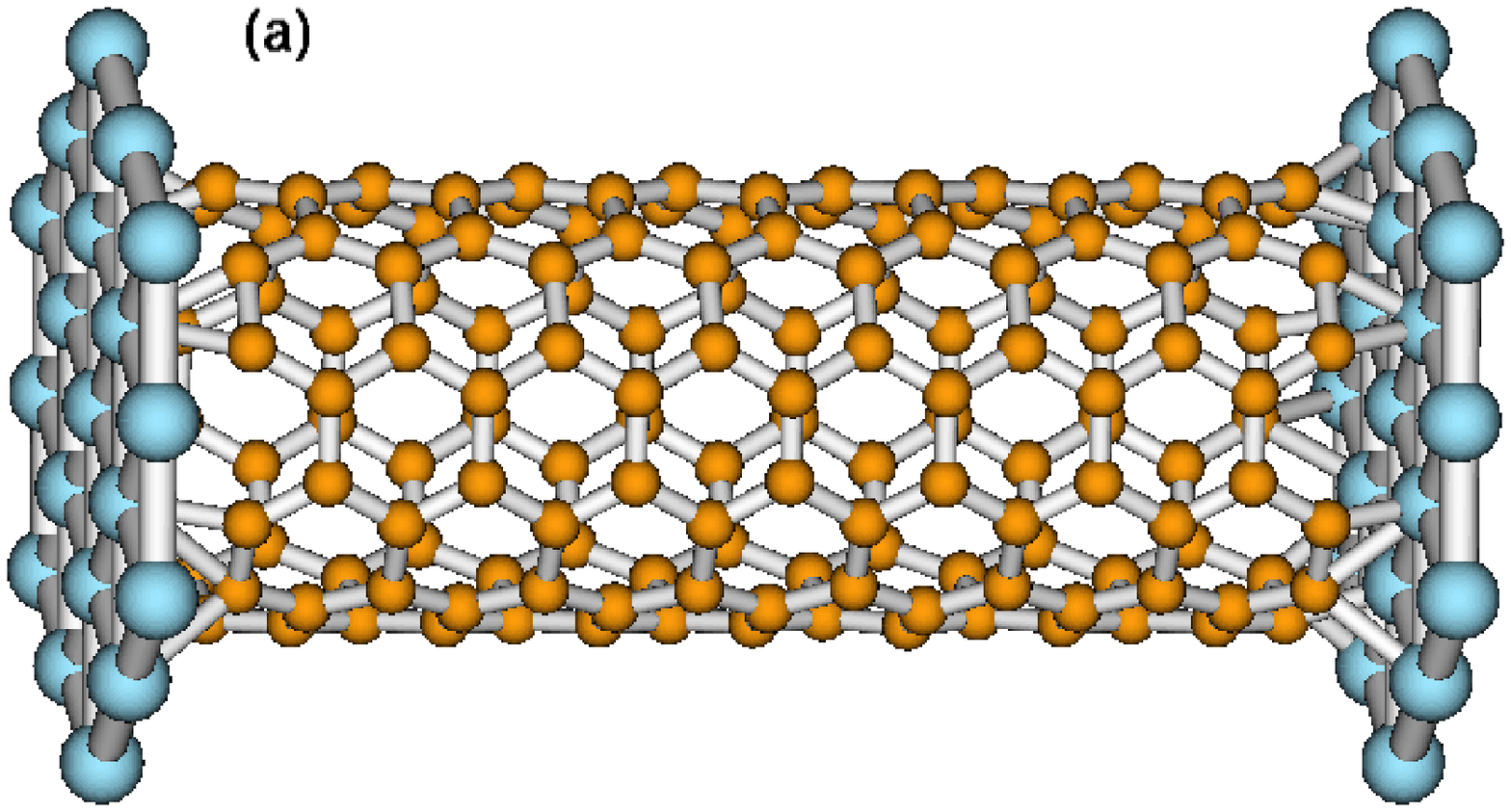}
\includegraphics[width=2.0in]{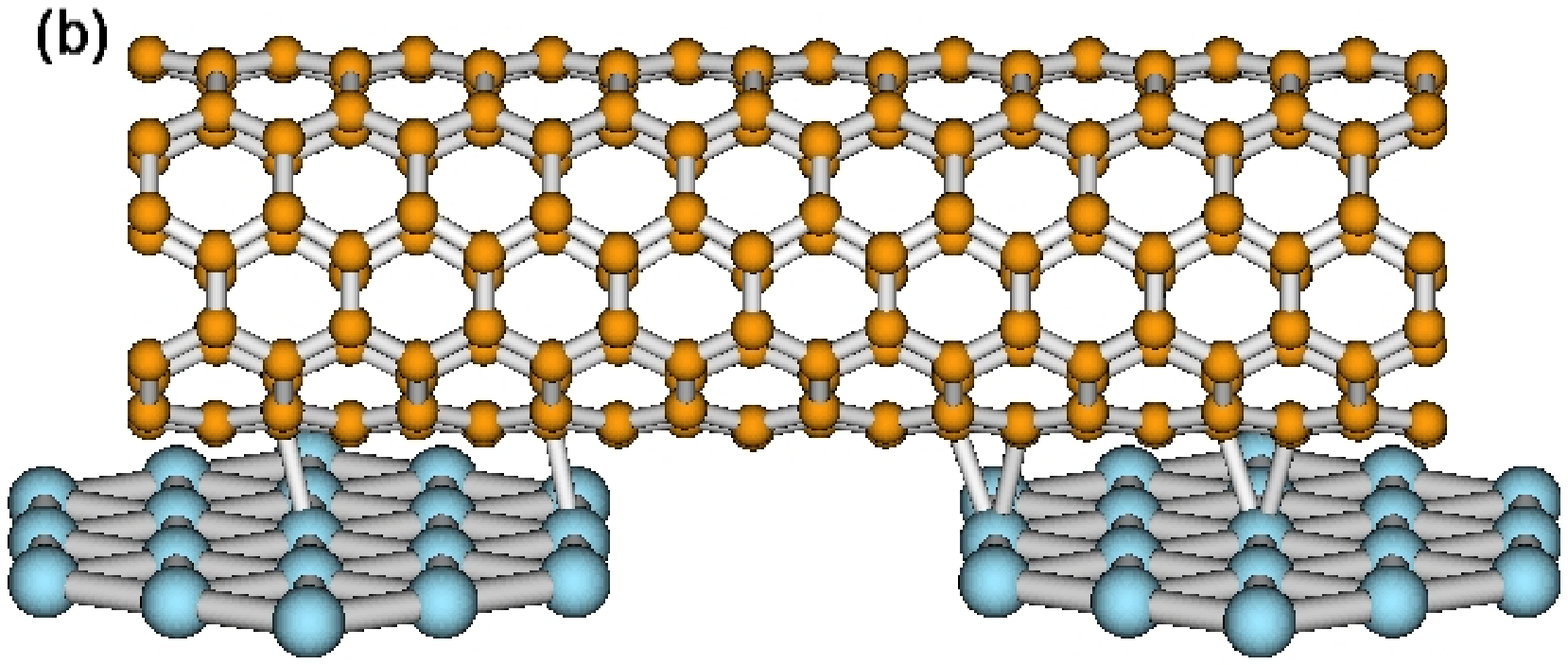}
\caption{ The two contact geometries considered in this work:
An open (5,5) carbon nanotube end-contacted to (111) surfaces (a) and  
the same nanotube side-contacted (b). \label{geom} }
\end{figure}

The actual atomic structure of the electrode (and probably that of the CNT)
at the contact are unknown and, most likely, 
change from sample to sample when fabricated under the same conditions.
Atomic-scale modeling, however, can still be of
guidance to the interpretation of the experiments and to the future design 
of operational devices with CNT's.
In this work we focus on the two key ingredients in this puzzle: 
The effect the atomic-scale geometry and the chemical 
nature of the electrode have on the transparency of the contact. We have
studied open single-walled metallic (5,5)  
CNT's contacted in two representative forms (see Fig.~\ref{geom}) 
to Al, Au, and Ti electrodes which 
are among the most commonly used metals in the experiments . 
From our {\em ab-initio} transport study  we 
find that in CNT's contacted to Al and Au electrodes
for end-contact geometry [see Fig.~\ref{geom}(a)]
the two CNT bands couple weakly to the electrodes. This allows us to 
resolve quasi-bound CNT states in the conductance and to estimate
the magnitude of the degeneracy removal due to Coulomb blockade effects 
in a direct manner. Moreover, we find that              
the two bands couple very differently to the electrodes 
(one of them is almost shut down for transport) and do not mix. For the 
side-contact geometry [see Fig.~\ref{geom}(b)] the coupling 
is the same for both bands, but similar in strength 
to the end-contact geometry.  Finally, our study presents the first direct
numerical evidence of
what has been hinted at on the basis of indirect first-principles
calculations\cite{Andriotis:apl:00,Yang:prb:02} and what has recently 
been observed in experiments\cite{Kong:prl:01}: 
Early 3-$d$ elements as Ti are probably the best choice for making 
high-transparency contacts to CNT's compared to more traditional metals such
as Al and Au. Although perfect transparency at
the contact is nerver achieved, our calculations indicate that
properly engineered Ti contacts are a good bet for future perfect 
contacts to CNT's.

From a theory point of view, the ``contact'' problem has been 
previously addressed\cite{Choi:prb:99,Anantram:apl:01,Derycke:apl:02},
but only partially. The reason is that a full analysis of this problem
 requires the use of sophisticated state-of-the-art
numerical techniques to calculate electrical transport
from first-principles\cite{Lang:prb:95,Yaliraki:jcp:98,Damle:prb:01}, 
where even the electrodes need to be described down to the atomic 
level\cite{Taylor:prb:01:a,Taylor:prb:01:b,Palacios:prb:01,Palacios:prb:02,Brandbyge:prb:02}.
These techniques are currently under development.
First of all, charge transfer at the contact, which aligns the chemical
potentials of the electrodes and the CNT, 
needs to be evaluated self-consistently\cite{Xue:prl:99,Rubio:prl:99}. 
Secondly, one needs to combine the {\em ab-initio}
calculation with Landauer's formalism\cite{Datta:book:95}.
Recently, we have presented a very promising approach, 
termed Gaussian Embedded Cluster Method\cite{Palacios:prb:01,Palacios:prb:02},
that allows us to address this problem in its full complexity.
Our method is based on standard quantum chemistry calculations performed with
the Gaussian98 code\cite{Gaussian:98}. 
A density functional (DF) calculation of a cluster comprising 
the CNT and a significant part of the electrodes is performed
(see Fig. \ref{geom}). Next, the retarded(advanced) Green's functions 
associated with the self-consistent hamiltonian or Fock operator $\hat{F}$ of
the cluster is modified to include the rest of the semi-infinite electrodes:
\begin{equation}
\left [(E\pm i\delta)-\hat F - \hat\Sigma^{(\pm)}
\right ] \hat G^{(\pm)}= \hat I.
\label{green}
\end{equation}
\noindent In this expression
$\hat\Sigma^{(\pm)}=\hat\Sigma_{\rm R}^{(\pm)} +
\hat\Sigma_{\rm L}^{(\pm)}$, where $\hat\Sigma_{\rm R}$($\hat\Sigma_{\rm L}$) 
denotes a self-energy operator
that accounts for the part of the right(left) semi-infinite electrode that has
not been included in the initial DF calculation\footnote{We choose
to describe the bulk electrode with a Bethe lattice tight-binding model
with the coordination and parameters appropriate for the 
electrodes\cite{Palacios:prb:01,Palacios:prb:02}. 
Details on the Bethe lattice parameters, the density functional, 
and the basis set used in our calculations
can be found in Ref.~\onlinecite{Palacios:prb:02}.},  and
$\hat I$ is the unity matrix. 
In a non-orthogonal basis, like those commonly used in Gaussian98,
the embedded cluster density matrix takes the form
\begin{equation}
P=-\frac{1}{\pi}\int_{-\infty}^{E_{\rm F}}{\rm Im}
\left[S^{-1} G^{(-)}(E) S^{-1} \right ]\; {\rm d}E,
\label{eqn:nab}
\end{equation}
where $S$ is the overlap matrix, $G^{(-)}$ is the retarded
Green's function expressed in the non-orthogonal basis, and
$E_{\rm F}$ is the Fermi energy which is set by imposing overall
charge neutrality in the cluster. The density matrix is returned to Gaussian98
and the process is repeated until the procedure 
converges.
The conductance can finally be
calculated through the standard expression\cite{Datta:book:95}:
\begin{equation}
{\mathcal G}=\frac{2e^2}{h}{\rm Tr}[T] = \frac{2e^2}{h}{\rm Tr}
[\Gamma_L G^{(-)}\Gamma_R G^{(+)}],
\label{g}
\end{equation}
where  Tr denotes the trace over all the orbitals in the cluster and 
where the matrices $\Gamma_R$ and $\Gamma_L$
are $i(\Sigma^{(-)}_R-\Sigma^{(+)}_R)$ and
$i(\hat\Sigma^{(-)}_L-\hat\Sigma^{(+)}_L)$, respectively. 
In order to single out the contribution of individual channels
to the current one can diagonalize the transmission matrix $T$.

\begin{figure}
\includegraphics[width=3.0in]{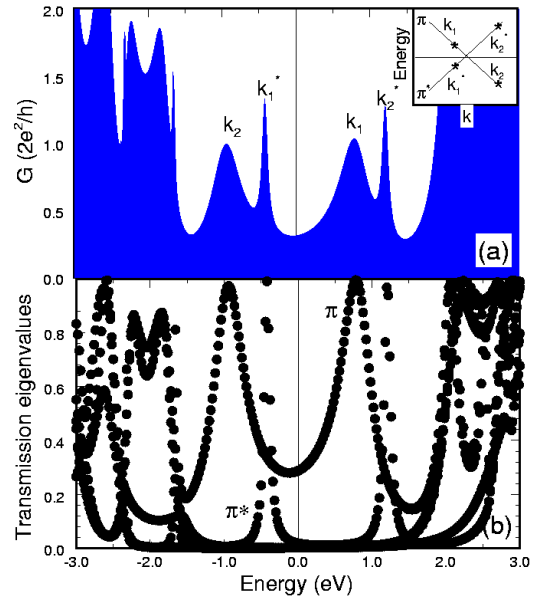}
\caption{ (a) Conductance as a function of energy for an $N=10$ (5,5) open
metallic nanotube end-contacted to a  Al(111) surface 
[see Fig.\ \ref{geom}(a)].  The nanotube-surface distance has been optimized 
to a value of 1.8 \AA and the Fermi energy is set to zero. 
Inset: Schematic band structure of the metallic
nanotube showing the four states responsible for the resonances.
(b) Transmission as 
a function of energy for the highest conducting channels. 
The symmetry of the two main channels is also shown.\label{end100} }
\end{figure}

Figure \ref{end100}(a) shows $\mathcal{G}$ around the Fermi energy
for a (5,5) metallic
CNT composed of $N=10$ carbon layers that has been end-contacted
[Fig.~\ref{geom}(a)] to  Al(111) surfaces (the end-carbon-layer--surface
distance has been optimized to a value of 1.8\AA)\footnote{A word of 
caution is due here.  Within DF theory only $\mathcal{G}(E_{\rm F})$
has a strict meaning. In order to obtain the zero-bias conductance
at different energies which would correspond to the conductance
for different values of an 
external gate potential which can charge or discharge the system,
one must perform the self-consistent calculation for 
a varying Fermi energy. We have 
analyzed the extent of this problem and found
that our conclusions are not modified significantly
as the charge in the system varies. This
partially justifies plotting
$\mathcal{G}(E)$ for neutral systems. However this problem might deserves
a further consideration when bound or
quasibound states are present in the CNT (see text below).}.
Four resonances appear around the Fermi energy (set to zero).
These resonances can be easily traced back to four
extended states of the isolated finite CNT\cite{Rubio:prl:99}.
Two of them ($k_1,k_2$) originate in
the bonding ($\pi$) band of the CNT and the other 
two ($k_1^*,k_2^*$) in the antibonding
($\pi^*$) band (see inset in Fig. \ref{end100}). The resonances
have different widths for different bands indicating that they couple 
very differently to the electrodes.
Moreover, the two bands do not mix with each other. This is more clearly seen in
Fig.~\ref{end100}(b) where we show the highest transmission eigenvalues of
the transmission matrix. Two independent channels exhibit resonances
in the energy window ($\approx 3.5 eV$) around $E_{\rm F}$
where only the $\pi$ and $\pi^*$ bands can contribute to transport. 
This result is consistent with the fact that $\pi^*$ states, of large angular
momentum, do not couple to the low-angular momentum states of the electrode,
while $\pi$ states, of low angular momentum, couple more 
easily\cite{Choi:prb:99,Anantram:apl:01}.  Notice that there is a charge
transfer from the metal to the CNT, but this
mainly localizes at the end carbon layer
($\approx 0.2$ per carbon atom) and it does not affect the overall
band positioning in the center of the CNT. 

\begin{figure}
\includegraphics[width=3.0in]{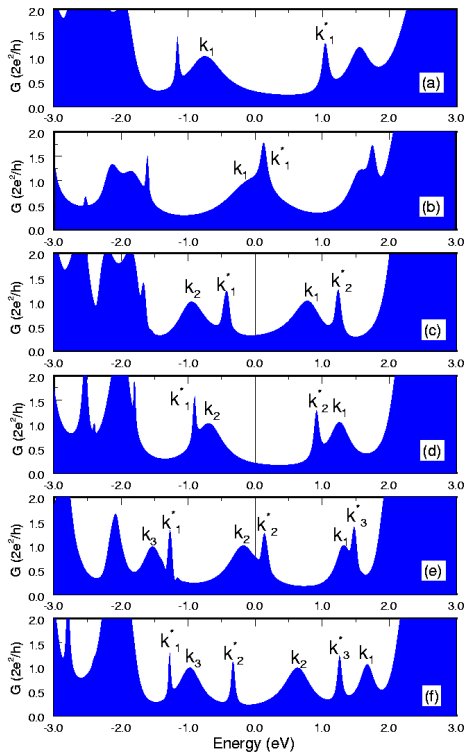}
\caption{Conductance as a function of energy for an $N=8$(a), $N=9$(b), 
$N=10$(c), $N=11$(d), $N=12$(e), and $N=13$(f) (5,5) open
metallic nanotube end-contacted to a  Al(111) surface 
[see Fig.\ \ref{geom}(a)].  The Fermi energy 
has been set to zero.\label{endX} }
\end{figure}

The specific band assignment of the resonances is nicely confirmed
by their evolution on the length of the CNT presented in 
Fig.~\ref{endX}. We have calculated the conductance
for $N=8,9,10,11,12$, and 13 carbon-layer CNT's. The opposite signs of the
group velocity for the $\pi$ and $\pi^*$ bands make the quasi-bound
states belonging to the $\pi^*$ band shift down in energies 
while those belonging to the $\pi$ band 
shift up as $N$ increases.  As expected from a
simple particle-in-a-box argument applied to finite
CNT's\cite{Rubio:prl:99}, for $N=3l$, where $l$ is an
integer, we should expect two states with the same wave vector $k_n$ but
in different bands to coincide at the Fermi energy. Naively one should thus
expect $\mathcal{G}=4e^2/h$\cite{Orlikowski:prb:01}.
Our results for the contacted $N=9$ and $N=12$ CNT's show otherwise: 
Two resonances never coincide at the Fermi level. The reason is that
Coloumb blockade prevents two (band and/or spin)
degenerate quasibound states to be filled up
at the same time and degeneracies are removed\footnote{We have
analyzed the Coulomb blockade process
in detail for the $N=9$ CNT. For a partially 
discharged CNT the two resonances labeled $k_1$ and $k_1^*$
coincide in energy above the 
Fermi energy and the conductance reaches there
4$e^2/h$. For the neutral [see Fig.~\ref{endX}(b)]
or slightly charged system this 
degeneracy is partially removed and the conductance drops.
The spin degeneracy removal due
to Coulomb blockade requires technically challenging 
open shell calculations and is currently under study.}. From 
Figs.~\ref{endX}(b) and (e)
we estimate the charging energy to be $\approx 0.3$ eV in these CNT's which
is smaller than the single-particle level spacing as confirmed by 
experiments\cite{Liang:prl:02}.

\begin{figure}
\includegraphics[width=3.0in]{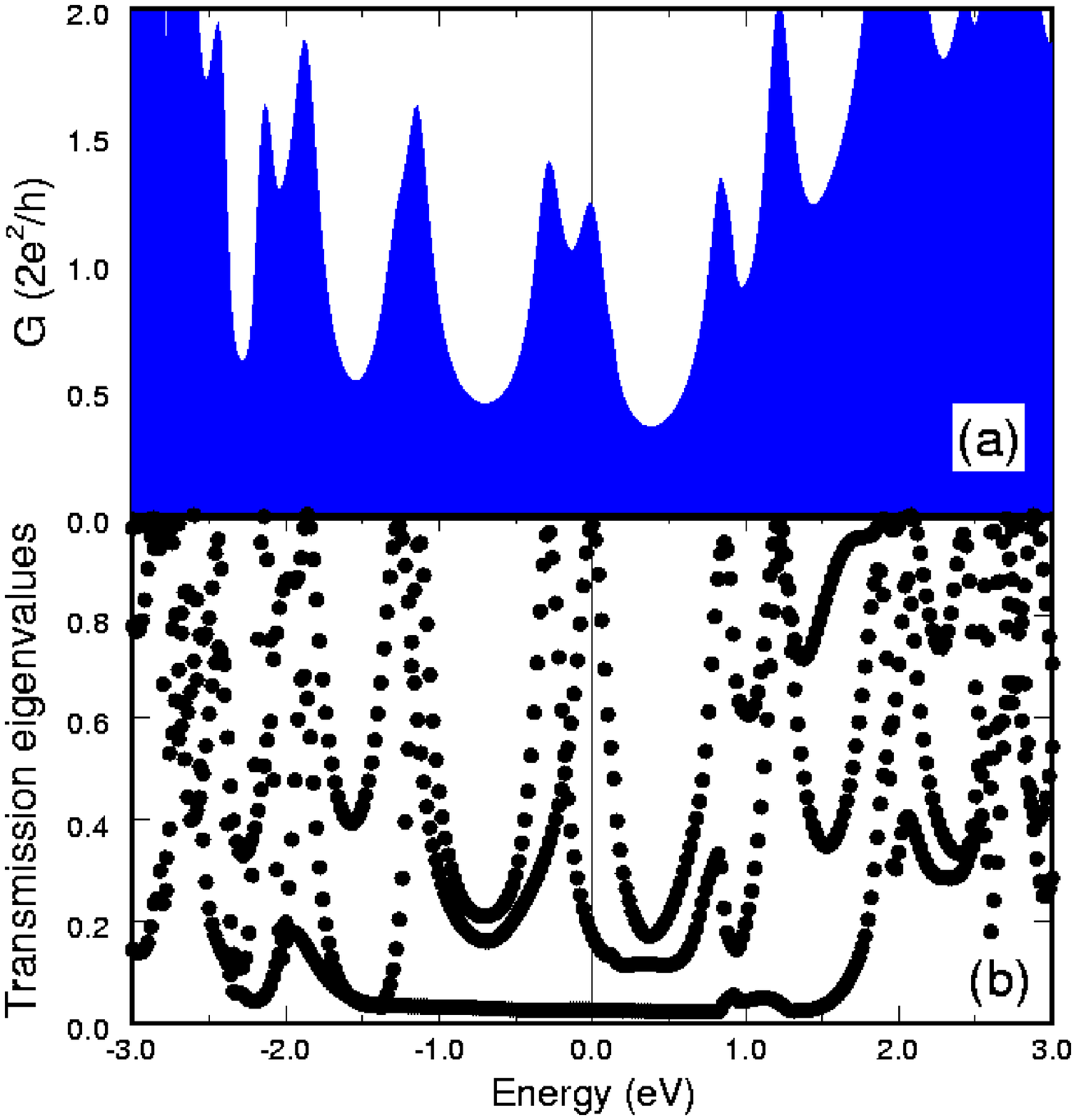}
\caption{ (a) Conductance as a function of energy for an $N=15$ (5,5) open
metallic nanotube side-contacted to a  Al(111) surface 
[see Fig.\ \ref{geom}(b)].  The nanotube-surface distance has been optimized 
to a value of 2.2 \AA.   (b) Transmission as 
a function of energy for the three highest conducting channels.\label{end150} }
\end{figure}

If the interpretation of the different coupling strengths of the
CNT bound states 
with the Al electrodes is correct and 
angular momentum considerations are relevant, similar couplings 
should be expected for both bands if no axial symmetry is present.
This is the case for the other contact geometry considered in this work [see
Fig.\ref{geom}(b)]. Figure  \ref{end150} shows results for an $N=15$
CNT side-contacted to  Al(111) surfaces (the CNT--surface distance has been
optimized to 2.3\AA). Conductance resonances come in pairs
in the relevant energy window which is what is
expected for an $N=15$ CNT.
More importantly, all of them present similar widths, 
confirming our expectations.  Contrary to the previous geometry, 
localized end states\cite{Rubio:prl:99} influence the 
coupling around 1eV for this contact geometry where mixing with the
CNT extended states takes place. 
Our results for the coupling strength with Al contacts are
consistent with previous studies where
jellium models were considered as contacts\cite{Anantram:apl:01},
and with those in Ref.~\onlinecite{Taylor:prb:01:a},
but we do not subscribe previous {\em ab-initio} results
presented in Ref.~\onlinecite{Nardelli:prb:01} based on what it seems to be
more realistic contact models similar to ours.  

\begin{figure}
\includegraphics[width=3.0in]{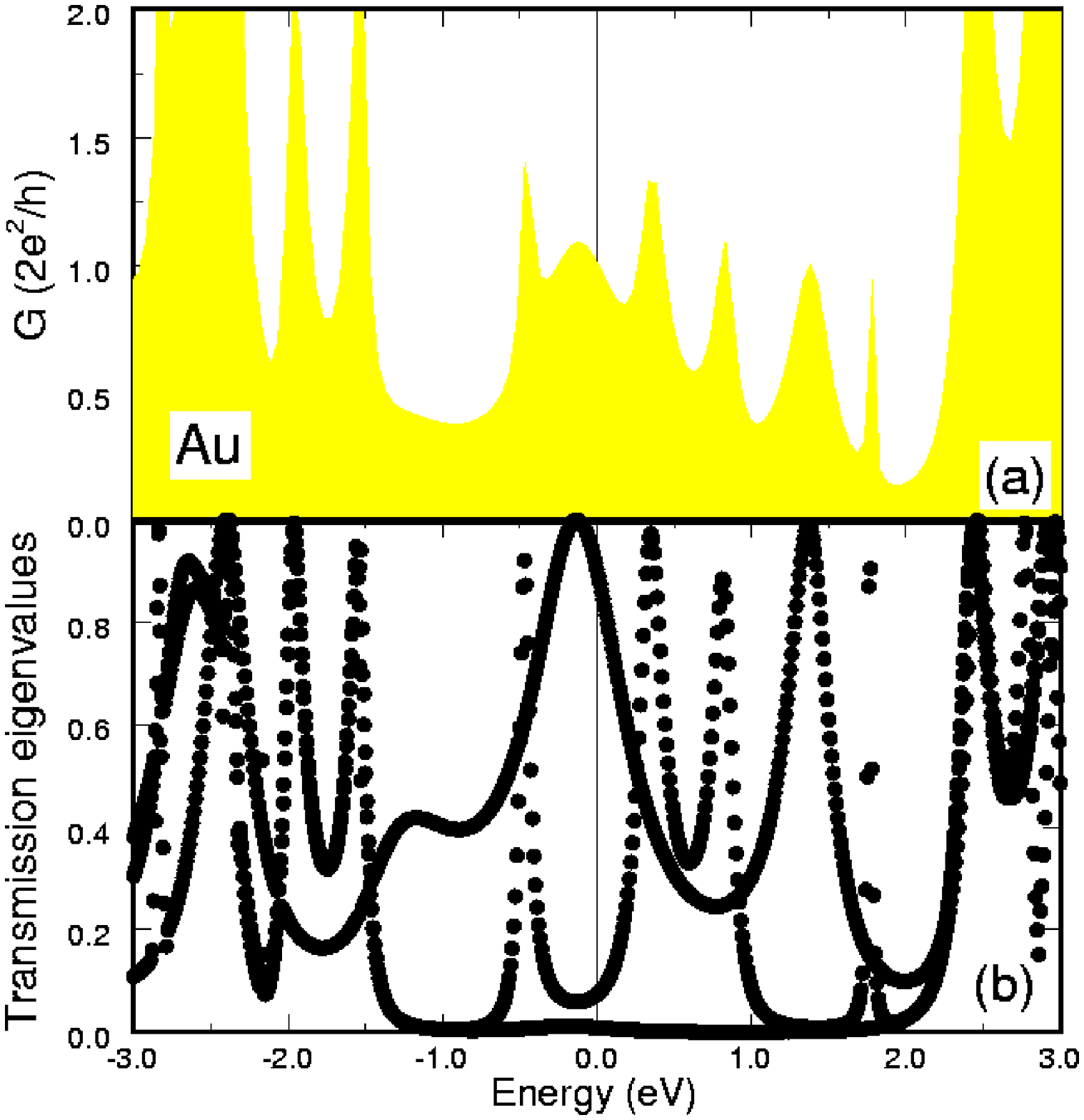}
\caption{ (a) Conductance as a function of energy for an $N=10$ (5,5) open
metallic nanotube end-contacted to a  Au(111) surface 
[see Fig.\ \ref{geom}(a)].  The nanotube-surface distance has been optimized 
to a value of 2.2 \AA.   (b) Transmission as 
a function of energy for the highest conducting channels.\label{end100Au} }
\end{figure}
\begin{figure}
\includegraphics[width=3.0in]{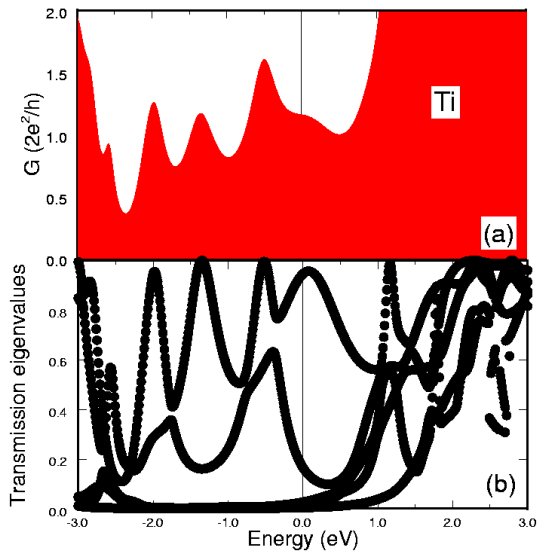}
\caption{(a) Conductance as a function of energy for an $N=10$ (5,5) open
metallic nanotube end-contacted to a  Ti(111) surface 
[see Fig.\ \ref{geom}(a)]. The nanotube-surface distance has been optimized 
to a value of 1.8 \AA.  (b) Transmission as a function of energy for the 
highest conducting channels.\label{end100Ti} }
\end{figure}

We now complete our study for end-contacted $N=10$ CNT's considering
Au and Ti electrodes (see Figs.~\ref{end100Au} and \ref{end100Ti}). 
Several resonances 
are clearly visible close to the Fermi energy for the case of Au, 
but, in contrast to 
Al electrodes, it is difficult to identify specific extended states as we
did above.
This is in part due to the mixing of the $\pi$ and $\pi^*$ bands
with the end states which, in addition,
induce extra channels in the conductance, although these
channels are only relevant for transport in very
short CNT's\footnote{A detailed analysis
of why the end states do not play a significant role for Al in 
end-contact geometries is deferred
for future work.}. Apart from this,
the coupling strength of the two bands is similar to that found for
Al electrodes despite of the fact that
the Mulliken population analysis reflects a
minor charge transfer from the electrode to the CNT.
In Fig.\ \ref{end100Au}(b) we appreciate that 
the $\pi$ band coupling is also stronger than 
that of the $\pi^*$ band. In contrast to
Al and Au electrodes, where $\mathcal{G}$ exhibits resonances, 
$\mathcal{G}$ presents an oscillatory behavior for Ti around $E_{\rm F}$.
This is accompanied, as the anticrossings in the transmission eigenvalues
reveal in Fig.~\ref{end100Ti}(b),
by band mixing. This result reflects, as suggested in Ref.
\onlinecite{Yang:prb:02},
that Ti couples differently to the CNT (due to the presence of $d$-states at 
the Fermi energy) and forms a better contact (the charge transfer is 
$\approx$ 0.4 electrons per C atom at the end layer). 
At this point, however, we 
can only speculate on the possibility of perfect transparency 
for other Ti electrode geometries.

We acknowledge support by the 
Spanish CICYT under Grant No. 1FD97-1358
and by the Generalitat Valenciana under Grants No. GV00-151-01
and GV00-095-2. J.J.P. thanks S. Y. Wu for encouraging this
work in its initial stages.

\bibliography{moletronics}

\end{document}